\documentclass[preprint,12pt]{elsarticle}

\usepackage{multicol}
\usepackage{graphicx}
\usepackage{booktabs}
\usepackage{amssymb,bm,mathrsfs,bbm,amscd}
\usepackage[tbtags]{amsmath}
\usepackage{lastpage}
\usepackage{verbatim}
\usepackage{comment}

\journal{}

\begin{document}

\begin{frontmatter}



\title{Investigation of antineutrino spectral anomaly with reactor simulation uncertainty \tnoteref{label1}}
 \tnotetext[label1]{Corresponding author}
\author{Xubo.Ma\corref{cor1}\fnref{label2}}
 \ead{maxb@ncepu.edu.cn}
\author[label2]{Jiayi.Liu}
\author[label2]{Jiayi.Xu}
\author[label2]{Fan.Lu}
\author[label2]{Yixue.Chen}
\address[label2]{North China Electric Power University, Beijing 102206, China \fnref{label2}}

\begin{abstract}
  Recently, three successful antineutrino experiments (Daya Bay, Double Chooz, and RENO) measured the neutrino mixing angle $\theta_{13}$; however, significant discrepancies were found, both in the absolute flux and spectral shape. Much effort has been expended investigating the possible reasons for the discrepancies. In this study, Monte Carlo-based sampling was used to evaluate the fission fraction uncertainties. We found that fission cross-section uncertainties are an important source of uncertainty for $^{235}$U, $^{239}$Pu, and $^{241}$Pu, but for $^{238}$U, elastic and inelastic cross-sections are more important. Among uncertainty related to manufacturing parameters, fuel density is the main uncertainty; however, the total manufacturing uncertainty was very small. The uncertainties induced by burnup were evaluated through the atomic density uncertainty of the four isotopes. The total fission fraction uncertainties from reactor simulation were 0.83\%, 2.24\%, 1.79\%, and 2.59\% for$^{235}$U, $^{238}$U, $^{239}$Pu, and $^{241}$Pu, respectively, at the middle of the fuel cycle. The total fission fraction uncertainty was smaller than the previously derived value of 5\%. These results are helpful for studying the reactor antineutrino anomaly and precisely measuring the antineutrino spectrum in the Daya Bay antineutrino experiment.
\end{abstract}


\end{frontmatter}

\section{Introduction}
Antineutrinos produced by nuclear reactors are used to study neutrino oscillations and search for signatures of nonstandard neutrino interactions in the kilometer-baseline reactor experiments Daya Bay\cite{Dayabay}, Double Chooz\cite{DoubleChooz}, and RENO\cite{RENO}. Antineutrinos can also be used to monitor reactor conditions to safeguard operations\cite{safe}. Although the neutrino mixing angle $\theta_{13}$ was successfully determined by these experiments, comparison of the measured spectra of kilometer- and short-baseline experiments to the most up-to-date predictions showed significant discrepancies both in the absolute flux and spectral shape. A 2.9$\sigma$ deviation was found in the measured inverse beta decay  positron energy spectrum compared to predictions. In particular, an excess of events at energies of 4 -- 6 MeV was found in the measured spectrum\cite{spc1Dayabay,spc2Dayabay,spc2Chooz}, with a local significance of 4.4$\sigma$. These results have brought home the notion that neutrino fluxes are not as well understood as had been thought. At present, it is not clear what physical processes give rise to the neutrino spectra bump. Much effort has been focused on the reactor antineutrino anomaly, which arose from improved calculations of the antineutrino spectra derived from a combination of information from nuclear databases with reference $\beta$ spectra\cite{a7,a8,a9,a10}.

In reactor antineutrino experiments, the following formula is usually applied to calculate the antineutrino spectrum for the reactor:
\begin{equation}
S(E_{\nu})=\frac{W_{th}}{\sum_{j}f_{j}e_{j}}\sum_{i}f_{i}S_{i}(E_{\nu}),
\label{flux_equation}
\end{equation}
where $W_{th} ($MeV/s$)$ is the reactor thermal power, $f_{j}$ is the fission fraction associated with each isotope ($^{235}$U, $^{238}$U, $^{239}$Pu, and $^{241}$Pu), $e_{j}$ is the thermal energy release per fission event for each isotope, and $S_{i}(E_{\nu})$ is a function of the $\bar{\nu}_{e}$ energy $E_{\nu}$ signifying the $\bar{\nu}_{e}$ yield per fission for each isotope. Therefore, the bump in the reactor neutrino spectra may be caused mainly by one isotope or by some important fission event\cite{spca1,spca2,spca3}.

 The antineutrino flux is an important source of uncertainty associated with measurements in reactor neutrino experiments. To evaluate uncertainties in reactor simulations, the calculated concentrations of each isotope using different reactor simulation codes were compared with a benchmark\cite{ZDjurcic2009}, and a proximate method was proposed to determine the fission fraction uncertainty using concentration uncertainty. Using the Takahama-3 benchmark and the largest burnup sample calculated with MURE and DRAGON, the concentration differences of each isotope between the calculated value and experimental data were about 5\%\cite{Jones}. The fission fractions for the Daya Bay reactor were also simulated using DRAGON and compared with results obtained with the SCIENCE code\cite{science}. It was found the average deviation for $^{235}$U, $^{238}$U, $^{239}$Pu, and $^{241}$Pu were 0.71\%, 4.2\%, 2.1\%, and 3.5\%\cite{mpla1ma}, respectively. However, the fission fraction uncertainty as a function of burnup was not reported. To investigate this question and determine the correlation coefficient of the fission fraction of different isotopes, a new Monte Carlo-based method was proposed using the one-group fission cross section and concentration of each isotope\cite{scima2}. That paper also discussed a coefficient to correlate the results with the burnup function. However, these previous studies did not solve the problem of identifying the main source of uncertainty in the fission fractions. Thus, the purpose of this study was to investigate fission fraction uncertainties induced by neutron cross section and fuel manufacturing uncertainties in a reactor simulation.

This paper is structured as follows. The Monte Carlo-based sampling method used for evaluation of uncertainties of fission fractions is introduced in Section \ref{mc}. To generate a library of samples for the transport calculation, the Sensitivity and Uncertainty Analysis Code for Light Water Reactor (SUACL) was developed, and the Three Mile Island Unit 1 (TMI-1) benchmark and a mixed oxide fuel (MOX) test were used to verify the code, as described in Section \ref{vf}. Section \ref{re} discusses how all the reaction types for the four isotopes and other important isotopes, atomic density uncertainties induced by burnup, and manufacturing parameters were taken into account to evaluate uncertainty of fission fractions. The last section provides conclusions.

\section{Monte Carlo-Based sampling method}
\label{mc}
In reactor simulations, we always focus on the neutronics characteristics because the fission power and chain reaction are determined by the neutron flux. Therefore, reactor simulation is generally called a neutronics calculation. The simulation results are affected by various uncertainties, including material composition, geometry, operation conditions, measured plant data, neutron cross section, and model approximations. In this study, the impact of cross section and fuel manufacturing parameter uncertainty on the fission fraction results was evaluated.

\begin{figure}
\begin{center}
\includegraphics[height=12cm,width=10cm,angle=-90]{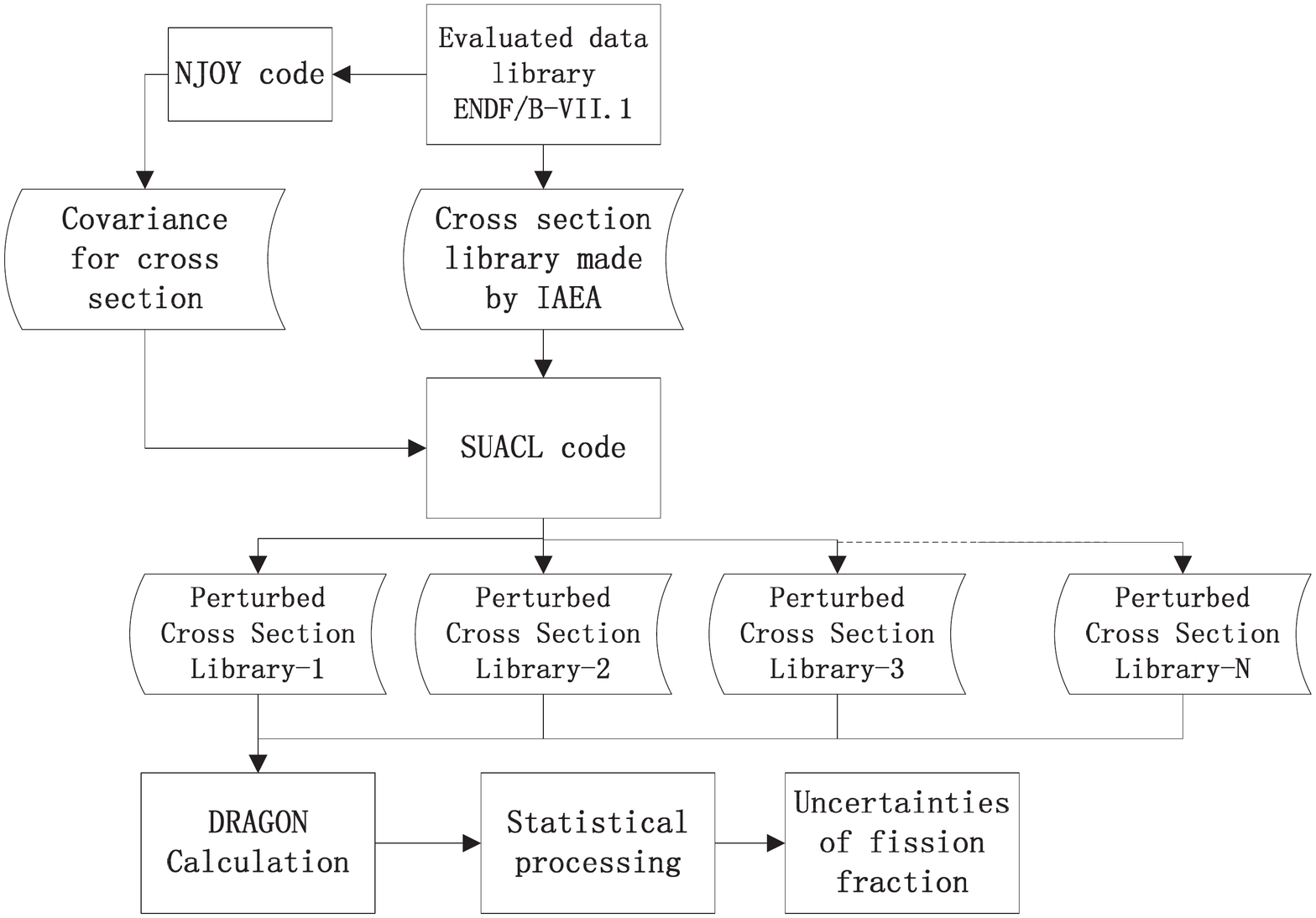}
\caption{Uncertainty estimation of fission fractions for antineutrino experiments using Monte Carlo-based sampling}
\label{flow1}
\end{center}
\end{figure}

\begin{figure}
\begin{center}
\includegraphics[width=14cm,height=7.5cm]{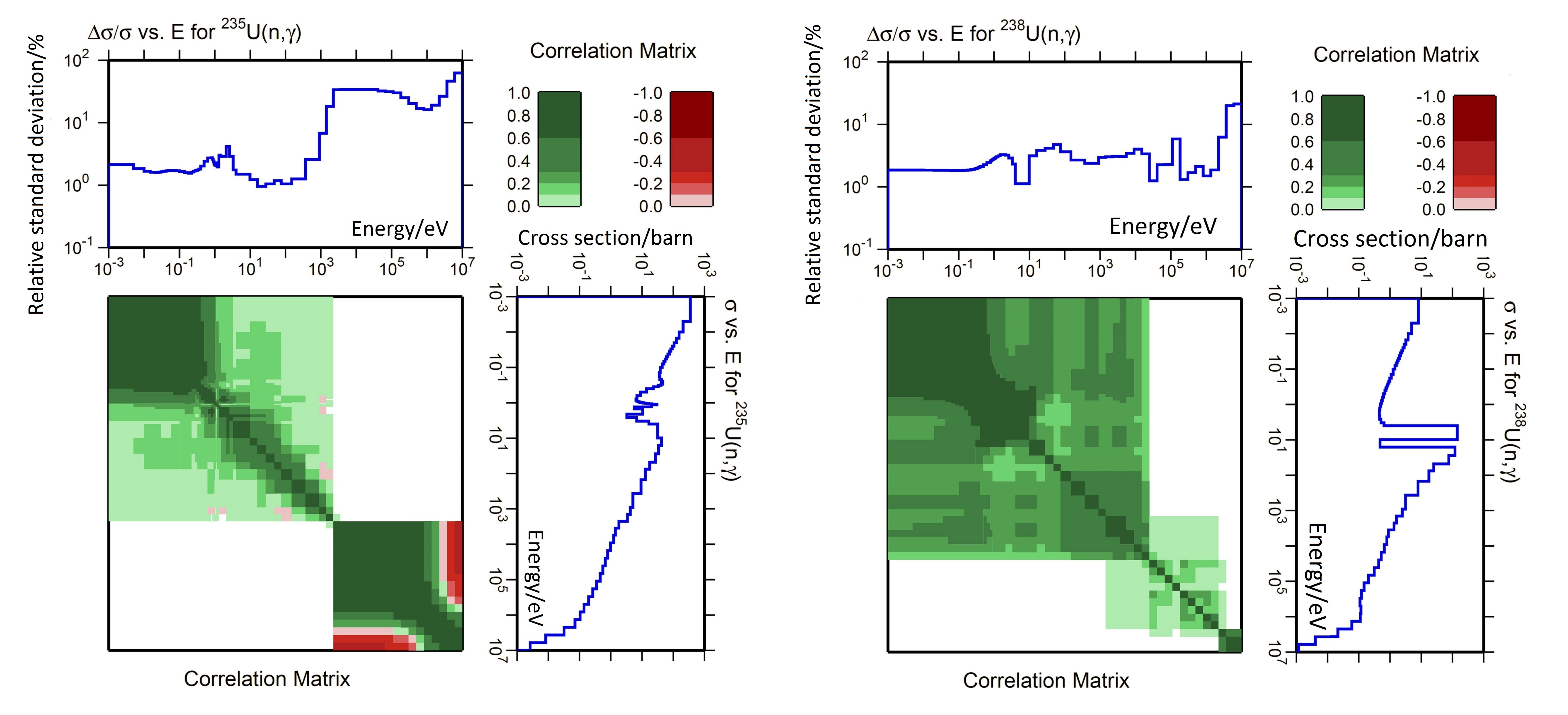}
\caption{The correlation matrix of $^{235}$U and $^{238}$U for (n,$\gamma$) cross section}
\label{covmatrix}
\end{center}
\end{figure}

Propagation of cross section uncertainties to core characteristics has been traditionally evaluated by the sandwich variance formula\cite{sandwich,sandwich1}. There are three difficulties in using the sandwich formula for pressurized water reactor (PWR) uncertainty analysis: complicated calculation sequence, nonlinear effects of thermal hydraulics and burnup, and the large number of input and output parameters. Recently, the Monte Carlo-based sampling method has attracted attention, since it can avoid these major difficulties. The procedures for uncertainty estimation of the fission fraction for antineutrino experiments using Monte Carlo-based sampling method is shown Fig.\ref{flow1}.
Generally, cross sections can be divided into basic and integral sections. The integral cross sections are composed of basic cross sections. The cross section uncertainties are stored in their covariance data. Different approaches to evaluating a nuclear data library may lead to different analysis results. The covariance data used in this study were produced using the NJOY code\cite{njoy} based on the ENDF/B-VII.1\cite{b7} and JENDL-4.0 libraries\cite{jendl}. The covariance matrices for $^{235}$U and $^{238}$U for the (n,$\gamma$) cross section are shown in Fig. \ref{covmatrix}. Other important inputs are the engineering parameters for fuel/assembly manufacturing\cite{benchmark}, which are crucial to the simulation model. The uncertainty of these parameters can propagate to the results and reduce their accuracy.
The SUACL code was designed specifically to analyze the uncertainty of neutron cross section and manufacturing parameters.

In the Monte Carlo-based sampling method, samples are randomly taken from the parent population. In this study, multi-group microscopic cross sections were randomly sampled using the covariance matrix. Multi-group means that the neutron energy have been divided many segments, and each segment is called one group. Sets of sampled cross sections approximately represent the distribution of the microscopic cross sections due to the nature of measurement error. Cross sections are randomly perturbed\cite{akioyamamoto} by

\begin{equation}
\sigma_{x,g}^{per}=P_{x,g}\times\sigma_{x,g}
\label{perequation}
\end{equation}
where $P_{x,g}$ is the perturbation factor for a microscopic cross section, $\sigma_{x,g}^{per}$ is the perturbed microscopic cross section, $\sigma_{x,g}$ is the unperturbed microscopic cross section in the original cross section library, $x$ is the type of cross section, and g is the energy group. The total sample number is set 100 for each reaction type of each isotope. Multi-group, microscopic cross sections in the original cross section library were perturbed by Eq. (\ref{perequation}) to generate the perturbed cross section library. A procedure for generating vectors of dependent random variables consistent with a given covariance matrix involves performing a spectral decomposition of the matrix.

\begin{equation}
\bf{\Sigma=V\times D\times V^{T}.}
\label{per2}
\end{equation}
The matrix of relative covariance,$\bf{\Sigma}$, is decomposed into three matrices, where {\bf{V}} is a matrix whose columns are eigenvectors of ${\bf{\Sigma}}$ and {\bf D} is a diagonal matrix of eigenvalues that correspond to the eigenvectors in {\bf V}.
Matrix $\bf{\Sigma^{1/2}}$ is defined as

\begin{equation}
\bf{\Sigma^{1/2}=V\times D^{1/2}\times V^{T}}
\label{per1}
\end{equation}
where {\bf $D^{1/2}$} is a diagonal matrix whose elements are the roots of the elements in {\bf D}.
The perturbed factor can be calculated using Eq. (\ref{perequation1})\cite{phdfull}.
\begin{equation}
\bf{P(\Sigma)=\Sigma^{1/2}G(0,1)+I},
\label{perequation1}
\end{equation}
where {\bf G} is a vector of $n$ normally distributed dependent random variables with a mean of zero and standard deviation of 1 and {\bf I} is a identity matrix.

\section{Validation of SUACL for PWR pin cell analysis}
\label{vf}
The SUACL code evaluates fission fraction uncertainty in the reactor simulation using Monte Carlo sampling\cite{suacl}. The perturbed nuclear library is generated by SUACL, as shown in Fig. \ref{flow1}, and then the reactor simulation code DRAGON\cite{dragon} applies the perturbed nuclear library to do the transport calculation. To verify the SUACL code, the TMI-1 PWR fuel cell benchmark\cite{benchmark} and MOX cell were used. Parameters of the TMI-1 and MOX datasets are shown in Table \ref{uo2mox}. The uncertainty of the infinity multiplication factors for the TMI-1 and MOX cells, which were evaluated using SUACL, are shown in Table \ref{umresult}. We found that the SUACL results were consistent with those obtained with other codes, such as TSUNAMI-1D, for the TMI-1 and MOX cells. 

\begin{table}
\begin{center}
\caption{\label{uo2mox}
Parameters of TMI-1 and MOX cell}
\footnotesize
\begin{tabular*}{100mm}{c@{\extracolsep{\fill}}ccc}
\toprule
Parameters & MOX cell & TMI-1 cell \\ \hline
Fuel material & (U,Pu)O$_{2}$ & UO$_{2}$ \\ \hline
Gap material & N/A & He gas \\ \hline
Clad material & Zircaloy-4 & Zircaloy-4 \\ \hline
Moderator     & H$_{2}$O  & H$_{2}$O   \\ \hline
Fuel pellet/mm &9.020  & 9.391          \\ \hline
Gap thickness/mm &0.0  & 0.955          \\ \hline
Clad thickness/mm & 0.380 & 0.673       \\ \hline
Unit cell pitch/mm & 12.60 & 14.427      \\
\bottomrule
\end{tabular*}
\end{center}
\end{table}

\begin{table}
\begin{center}
\caption{\label{umresult}
Uncertainty of infinity multiplication factor k$_{inf}$ (\%)}
\footnotesize
\begin{tabular*}{145mm}{c@{\extracolsep{\fill}}cccccc}
\toprule
Isotopes~~  & Reaction Parameters & CASMO-4\cite{casmo4} & TSUNAMI\cite{TSUNAMI} & UNICORN\cite{unicorn} & SUACL \\ \hline
UO$_{2}$ & & & & & \\ \hline
$^{238}$U & $\sigma_{n,\gamma}$ &0.325 & 0.284 & 0.377& 0.394 \\ \hline
$^{238}$U & $\sigma_{elastic}$ &0.039 & 0.106 &0.114 &0.211 \\ \hline
$^{235}$U & $\sigma_{n,\gamma}$ &0.223 &0.210&0.196&0.209\\ \hline
$^{235}$U &
$\sigma_{f}$ &0.078 & 0.077 &0.079 &0.084 \\ \hline
H &
$\sigma_{elastic}$ &0.027&0.026& 0.038&0.037\\ \hline

MOX & & & & & \\ \hline
$^{239}$Pu & $\sigma_{n,\gamma}$ & N/A & 0.204& N/A&0.211\\ \hline
$^{239}$Pu & $\sigma_{f}$ &N/A & 0.192 & N/A & 0.182 \\ \hline
$^{240}$Pu & $\sigma_{n,\gamma}$ &N/A & 0.088&N/A&0.063\\ \hline
$^{242}$Pu & $\sigma_{n,\gamma}$ &N/A &0.012&N/A&0.011\\ \hline
$^{238}$U & $\sigma_{n,\gamma}$ &N/A & 0.215&N/A&0.284\\ \hline
$^{238}$U & $\sigma_{f}$ &N/A&0.018&N/A& 0.017 \\ \hline
$^{235}$U & $\sigma_{n,\gamma}$ &N/A&0.061&N/A&0.059\\ \hline
$^{235}$U & $\sigma_{f}$ &N/A & 0.025& N/A & 0.020\\ \hline
H & $\sigma_{elastic}$ &N/A & 0.038 & N/A & 0.030 \\

\bottomrule
\end{tabular*}
\end{center}
\footnotesize
{$\sigma_{n,\gamma}$ represent capture cross section,$\sigma_{f}$ fission cross section, $\sigma_{elastic}$  elastic scattering cross section}

\end{table}

\section{Reactor simulation uncertainty for antineutrino experiment}
\label{re}
Four isotopes, namely $^{235}$U, $^{238}$U,$^{239}$Pu, and $^{241}$Pu, are important to antineutrino experiments because more than 99.0\% of the antineutrinos are emitted from these isotopes. To predict antineutrino flux for an antineutrino experiment, the fission fraction of each isotope is needed according to Eq. (\ref{flux_equation}). The process of evaluating the fission fraction depends on how neutrons are handled in the reactor simulation . In general, the neutron behavior is described by a neutron transport equation and the cross sections of neutrons with matter are the coefficients of the transport equation. The fission fraction of isotopes $f_{i}$ ($i$=$^{235}$U, $^{238}$U,$^{239}$Pu, and $^{241}$Pu) can be defined as
\begin{equation}
f_{i}=\frac{N_{i}\sum^{G}_{g=1}\sigma_{i,g,f}\phi_{g}}{\sum_{i}N_{i}\sum_{g=1}^{G}\sigma_{i,g,f}\phi_{g}},
\label{ff}
\end{equation}
where $\sigma_{i,g,f}$ is the microscopic cross section of isotope $i$ in group g, g is the neutron energy number, G is the total neutron energy group number, $N_{i}$ is the atomic density of the isotope, and $\phi_{g}$ is the neutron flux of group $g$ (as calculated by DRAGON). From Eq. (\ref{ff}), we can see that the fission fraction uncertainties can be divided into three parts: (1) atomic density uncertainty caused by burnup, (2) fission cross section uncertainty, and (3) neutron flux uncertainty caused by other cross sections and parameters.

\begin{figure}
\begin{center}
\includegraphics[width=10cm]{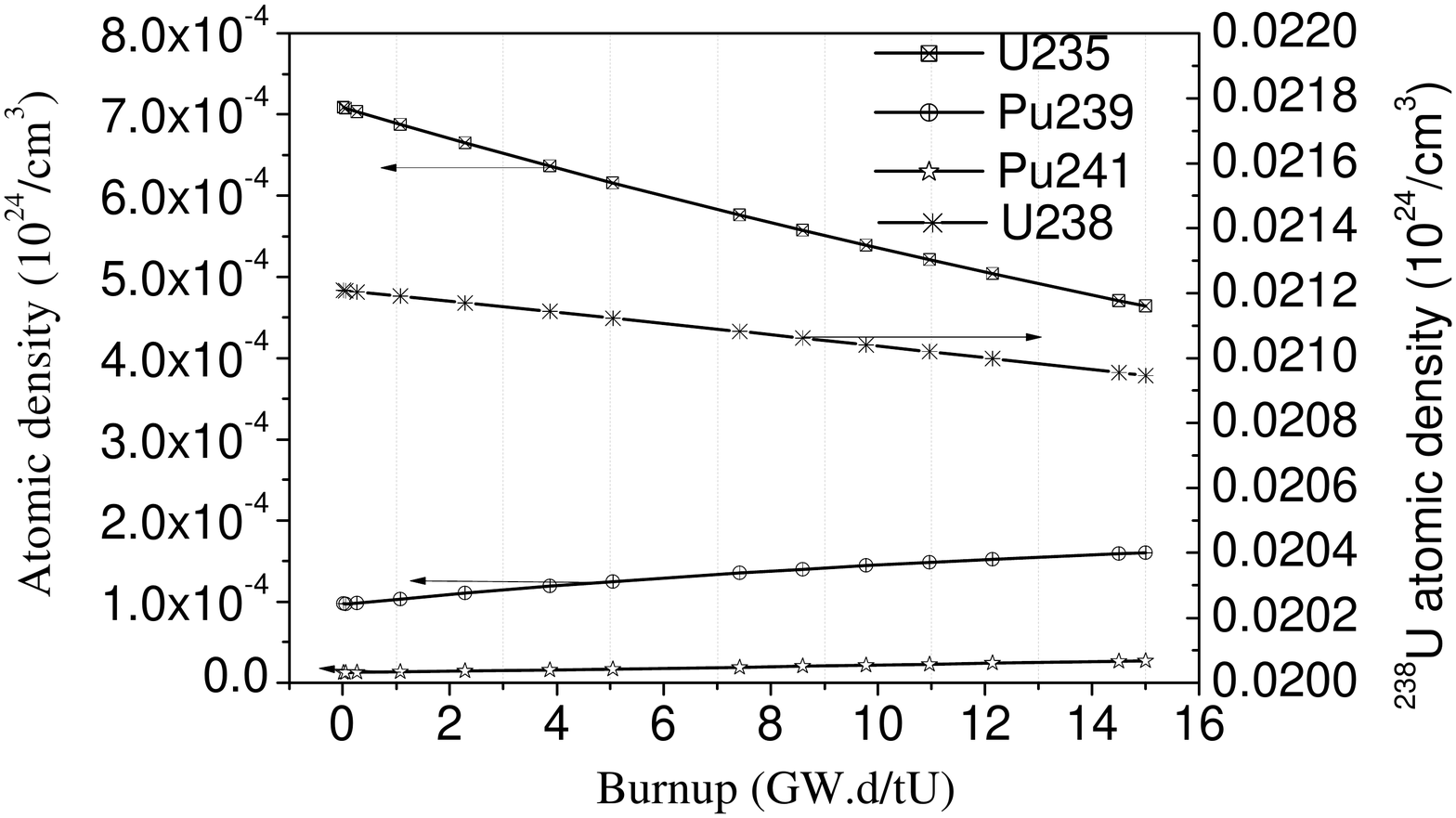}
\caption{Atomic density as a function of burnup}
\label{atomicdensiry}
\end{center}
\end{figure}

\begin{figure}
\begin{center}
\includegraphics[width=10cm]{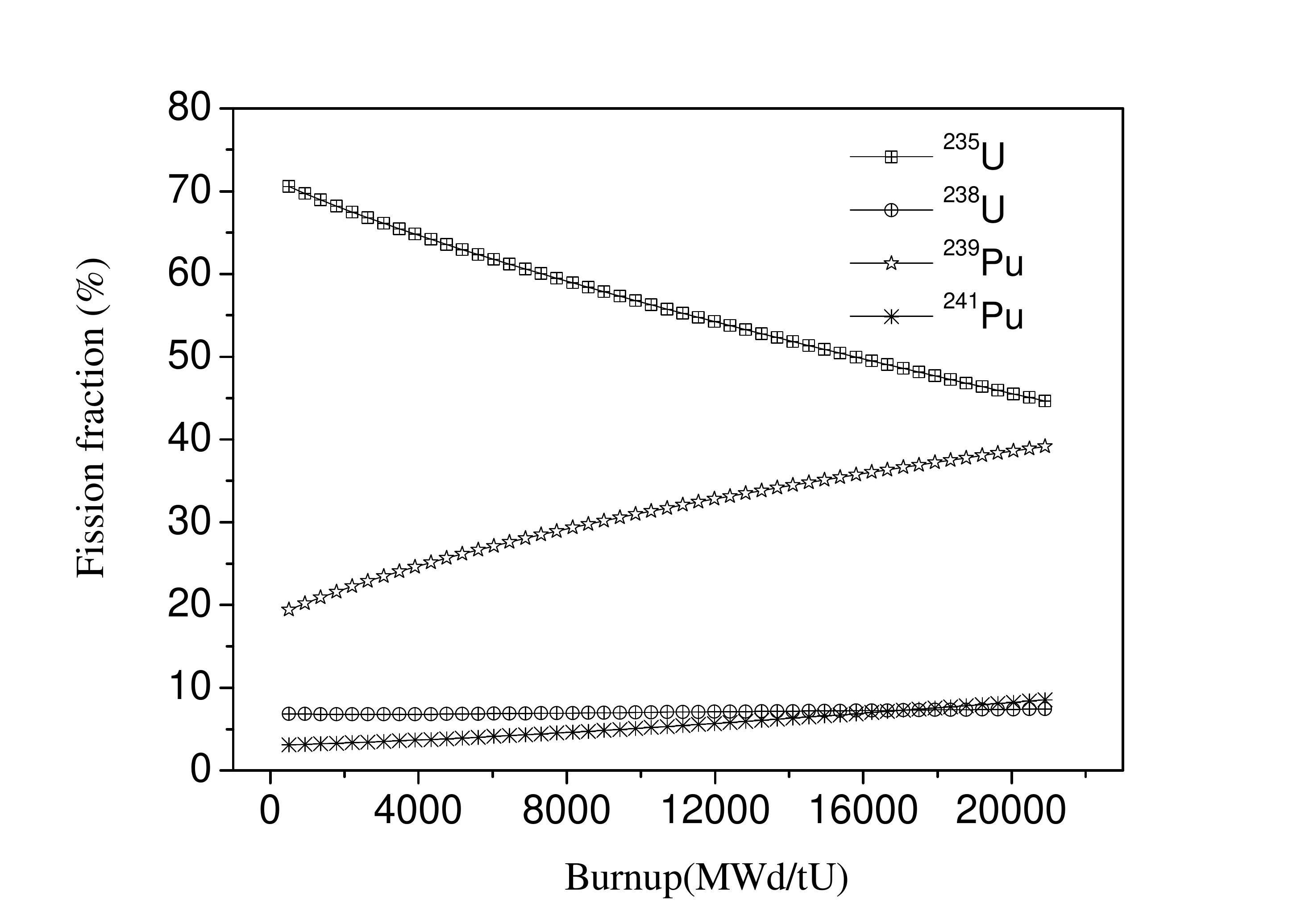}
\caption{Fission fraction of $^{235}$U,$^{238}$U,$^{239}$Pu and $^{241}$Pu as a function of burnup calculated by DRAGON}
\label{fissionfraction}
\end{center}
\end{figure}

The Daya Bay reactor operates with 157 fuel assemblies to produce a total thermal power of 2,895 MW. The assembly is a 17 $\times$ 17 design, for a total of 289 rods, including 264 fuel rods, 24 control rods, and 1 guide tube. The enrichment of fuel for the Daya Bay reactor core is 4.45\% for an 18-month reload interval\cite{mw900}. About one-third of the fuel in the core is reloaded. The atomic density of the four isotopes as a function of burnup is shown in Fig. \ref{atomicdensiry}, and the fission fraction as a function of burnup is shown in Fig. \ref{fissionfraction}. The fission fraction of $^{235}$U decreased with increasing burnup mainly because the atomic density decreased. However, $^{239}$Pu had the opposite trend, mainly because its atomic density increased.

\begin{figure}
\begin{center}
\includegraphics[width=9cm]{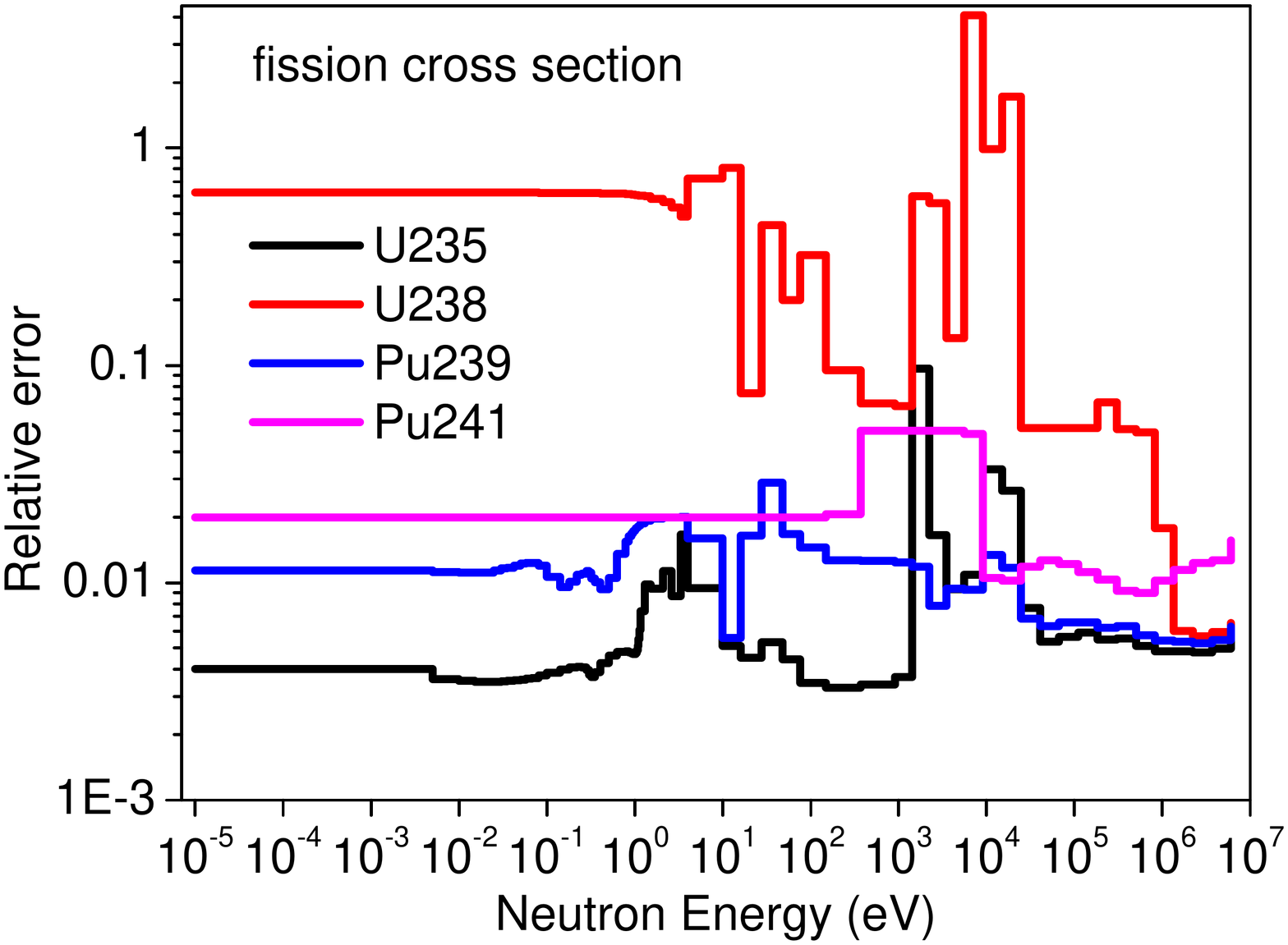}
\caption{fission cross section relative error vs the neutron energy }
\label{fissioner}
\end{center}
\end{figure}

The cross sections are one of main sources of uncertainty in evaluating the fission fraction. The most important reaction types for each isotope, such as fission cross section $\sigma_{f}$, capture cross section $\sigma_{n,\gamma}$, elastic cross section $\sigma_{elas}$, and inelastic cross section $\sigma_{inelas}$, as well as neutrons produced  per fission $\nu$, were accounted for. The fission fraction uncertainties of each isotope caused by different reaction types at the beginning of cycle (BOC), middle of cycle (MOC), and end of cycle (EOC) are given in Table \ref{um1}, Table \ref{um2}, and Table \ref{um3}, respectively. When we evaluated the fission fraction uncertainties at the MOC and EOC, the uncertainties of cross section were taken into account and the uncertainties of atomic density were ignored. Instead, the uncertainties of atomic density due to burnup were taken into account as described below . Two major conclusions can be drawn from the results given in the tables. First, the fission cross section is the main contributor to the fission fraction uncertainties because its uncertainty directly propagates to the fission fraction, and other cross section uncertainties indirectly propagate to the fission fraction through neutron flux uncertainties. Second, the fission fraction uncertainties of $^{235}$U, $^{239}$Pu, and $^{241}$Pu are induced by thermal neutrons because they are fissile and have large fission cross sections at low energy. However, the uncertainty for $^{238}$U is induced by fast neutrons because it is fissionable but only when the incident neutron energy is larger than the fission threshold energy. The relative errors of the fission cross section are shown in Fig. \ref{fissioner}. Although the relative error of $^{238}$U at low energy is very large, the relative error at high energy is relatively small. Therefore, the overall fission fraction uncertainty of $^{238}$U is not very large. The fission fraction uncertainty of $^{238}$U is dominated by the elastic and inelastic cross sections, as shown in Fig. \ref{fissionfraction}, because of large relative error in the high-energy region for elastic and inelastic cross sections.

The atomic density of the four isotopes varies because of transmutation and fission during reactor operation. Atomic density uncertainties come from the uncertainties of decay data, fission yields, and neutron flux. The results shown in Table \ref{um1}, Table \ref{um2}, and Table \ref{um3} neglect atomic number density uncertainties. Atomic density uncertainties as a function of burnup when propagating incident neutron data uncertainties according to ENDF/B-VII.1 covariance data were studied previously\cite{cjd}. Atomic density uncertainties are important at the EOC (60GW.d/tU), and the largest uncertainties of $^{235}$U, $^{239}$Pu, and $^{241}$Pu were 1.78\%, 2.0\%, and 1.91\%, respectively. The uncertainty of $^{238}$U was less than 1.0\%. To conservatively evaluate the fission fraction uncertainties induced by atomic density, the uncertainty of $^{238}$U was assumed as 1.0\%. The fission fraction uncertainties induced by atomic density as a function of burnup are shown in Table \ref{umb}. The values were 0.75\%, 0.846\%, 1.68\%, and 2.14\% for $^{235}$U, $^{238}$U, $^{239}$Pu, and $^{241}$Pu, respectively .

Fuel and assembly manufacturing uncertainties, such as enrichment, pellet density, cladding dimensions, burnable poison concentration, and assembly geometry, are also sources of uncertainty in reactor simulation. Manufacturing uncertainties in terms of $3\sigma$ are provided in Ref. \cite{benchmark}. Fission fraction uncertainties caused by manufacturing parameters in terms of $1\sigma$ are shown in Table \ref{umm}. We found that fuel density uncertainty is the main uncertainty among all the manufacturing parameters, and the other manufacturing parameter uncertainties are small.

Considering the above analysis, the total uncertainties of fission fractions in reactor simulation are shown in Table \ref{umt}. It can be seen that the total fission fraction uncertainties are 0.83\%, 2.24\%, 1.79\%, and 2.58\% for$^{235}$U, $^{238}$U, $^{239}$Pu, and $^{241}$Pu at the middle of cycle, respectively.

\begin{table}
\begin{center}
\caption{\label{um1}
Relative fission fraction uncertainties induced by different reaction type at the Begin Of Cycle (BOC)(\%)}
\footnotesize
\begin{tabular*}{140mm}{c@{\extracolsep{\fill}}cccccc}
\toprule
Isotopes~~&Reaction Channel &$^{239}$Pu & $^{241}$Pu & $^{235}$U &$^{238}$U\\ \hline
$^{235}$U &$\sigma_{f}$ &2.37E-1&2.47E-1 &1.22E-1&6.59E-2 \\ \hline
  & $\sigma_{n,\gamma}$ &7.68E-3&4.68E-3&7.66E-3&7.16E-2\\\hline
  & $\sigma_{inelas}$ &3.69E-4 &3.79E-4&3.66E-4&5.23E-3\\ \hline
  & $\sigma_{elas}$ &5.74E-4&5.83E-4&5.65E-4&8.06E-3\\ \hline
& $\nu$ &1.66E-2 &1.46E-02&1.07E-2&1.81E-1\\ \hline
$^{238}$U &$\sigma_{f}$ &3.14E-2&3.13E-2&3.11E-2&4.62E-1\\ \hline
  & $\sigma_{n,\gamma}$ &1.44E-2&2.15E-2&1.51E-2&2.10E-1\\ \hline
  & $\sigma_{inelas}$ &6.79E-2&7.02E-2&6.79E-2&9.94E-1\\ \hline
  & $\sigma_{elas}$ &9.65E-2&1.00E-1&9.54E-2&1.42E-0\\ \hline
  & $\nu$ &2.83E-3&2.49E-3&1.82E-3&3.09E-2 \\ \hline
$^{239}$Pu &$\sigma_{f}$ &5.91E-1&2.42E-1&2.32E-1&4.26E-2 \\ \hline
  & $\sigma_{n,\gamma}$ &3.04E-2&9.44E-3&8.52E-3&6.41E-2\\ \hline
  & $\sigma_{inelas}$ &1.02E-4&1.05E-4&1.02E-4&1.44E-3\\ \hline
  & $\sigma_{elas}$ &1.70E-4&1.74E-4&1.68E-4&2.40E-3\\ \hline
  & $\nu$ &1.75E-3&1.54E-3&1.13E-3&1.91E-3\\ \hline
$^{241}$Pu &$\sigma_{f}$ &4.91E-2&1.42E0&4.84E-2&9.73E-3 \\ \hline
  & $\sigma_{n,\gamma}$ &1.12E-3 &1.38E-3&4.68E-4&9.82E-3\\ \hline
  & $\sigma_{inelas}$ &2.87E-5&2.85E-5&2.84E-5&3.88E-4\\ \hline
  & $\sigma_{elas}$ &1.74E-5&2.28E-5&1.78E-5&2.17E-4\\ \hline
  & $\nu$ &7.77E-4&6.86E-4&5.01E-4&8.50E-3\\ \hline
$^{16}$O & $\sigma_{inelas}$ &5.45E-5 &5.41E-5&5.48E-5&7.83E-4\\ \hline
  & $\sigma_{elas}$ &7.15E-2&3.46E-2&1.62E-2&4.42E-1\\ \hline
H & $\sigma_{elas}$ &5.45E-2&6.72E-2&6.18E-2&8.92E-1\\ \hline
Total & & 6.58E-01	& 1.47E+00& 3.00E-1&2.08E+0 \\

\bottomrule

\end{tabular*}
\end{center}
\footnotesize
{$\sigma_{n,\gamma}$ represent capture cross section,$\sigma_{f}$ fission cross section, $\sigma_{elas}$  elastic scattering cross section,$\sigma_{inelas}$  inelastic scattering cross section, $\nu$ neutron number per fission}
\end{table}

\begin{table}
\begin{center}
\caption{\label{um2}
Relative fission fraction uncertainties caused by different reaction type at the Middle Of Cycle (MOC)(\%)}
\footnotesize
\begin{tabular*}{140mm}{c@{\extracolsep{\fill}}cccccc}
\toprule
Isotopes~~&Reaction Channel &$^{239}$Pu & $^{241}$Pu & $^{235}$U &$^{238}$U\\ \hline
$^{235}$U &$\sigma_{f}$ &2.14E-1&2.22E-1 &1.45E-1&6.06E-2 \\ \hline
  & $\sigma_{n,\gamma}$ &5.21E-3&4.22E-3&7.47E-3&5.87E-2\\\hline
  & $\sigma_{inelas}$ &3.70E-4 &3.84E-4&3.70E-4&5.36E-3\\ \hline
  & $\sigma_{elas}$ &5.76E-4&5.90E-4&5.73E-4&8.28E-3\\ \hline
  & $\nu$ &1.32E-2&1.16E-2&8.48E-3&1.49E-1\\ \hline
$^{238}$U &$\sigma_{f}$ &3.20E-2&3.19E-2&3.17E-2&4.61E-1\\ \hline
  & $\sigma_{n,\gamma}$ &1.41E-2&2.17E-2&1.61E-2&2.12E-1\\ \hline
  & $\sigma_{inelas}$ &6.91E-2&7.15E-2&6.84E-2&9.94E-1\\ \hline
  & $\sigma_{elas}$ &9.82E-2&1.02E-1&9.71E-2&1.42E-0\\ \hline
  & $\nu$ &2.72E-3&2.40E-3&1.75E-3&3.07E-2 \\ \hline
$^{239}$Pu &$\sigma_{f}$ &5.41E-1&2.85E-1&2.71E-1&5.13E-2 \\ \hline
  & $\sigma_{n,\gamma}$ &3.57E-2&9.13E-3&1.39E-2&7.61E-2\\ \hline
  & $\sigma_{inelas}$ &1.35E-4&1.41E-4&1.35E-4&1.96E-3\\ \hline
  & $\sigma_{elas}$ &2.23E-4&2.35E-4&2.67E-4&3.28E-3\\ \hline
  & $\nu$ &2.30E-3&2.03E-3&1.48E-3&2.60E-2\\ \hline
$^{241}$Pu &$\sigma_{f}$ &7.58E-2&1.40E0&7.51E-2&1.65E-2 \\ \hline
  & $\sigma_{n,\gamma}$ &1.60E-3&1.92E-3&4.58E-4&1.33E-2\\ \hline
  & $\sigma_{inelas}$ &4.13E-5&4.17E-5&4.10E-5&5.79E-4\\ \hline
  & $\sigma_{elas}$ &2.63E-5&3.38E-5&2.30E-5&3.12E-4\\ \hline
  & $\nu$ &1.14E-3&1.01E-3&7.36E-4&1.29E-2\\ \hline
$^{16}O$ & $\sigma_{inelas}$ &5.49E-5&5.80E-5&5.97E-5&8.22E-4\\ \hline
  & $\sigma_{elas}$ &6.94E-2&2.95E-2&9.74E-3&4.36E-1\\ \hline
H & $\sigma_{elas}$ &5.85E-2&6.95E-2&6.27E-2&8.95E-1\\ \hline
Total & & 6.08E-1& 1.45E+0&	3.46E-1	&2.07E+0 \\

\bottomrule

\end{tabular*}
\end{center}
\footnotesize
{$\sigma_{n,\gamma}$ represent capture cross section,$\sigma_{f}$ fission cross section, $\sigma_{elas}$  elastic scattering cross section,$\sigma_{inelas}$  inelastic scattering cross section, $\nu$ neutron number per fission}
\end{table}

\begin{table}
\begin{center}
\caption{\label{um3}
Relative fission fraction uncertainties caused by different reaction type at the End Of Cycle (EOC)(\%)}
\footnotesize
\begin{tabular*}{140mm}{c@{\extracolsep{\fill}}cccccc}
\toprule
Isotopes~~&Reaction Channel &$^{239}$Pu & $^{241}$Pu & $^{235}$U &$^{238}$U\\ \hline
$^{235}$U &$\sigma_{f}$ &1.81E-1&1.87E-1 &1.80E-1&5.33E-2 \\ \hline
  & $\sigma_{n,\gamma}$ &3.60E-3&3.87E-3&6.97E-3&4.81E-3\\\hline
  & $\sigma_{inelas}$ &3.02E-4&3.14E-4&3.01E-4&4.34E-4\\ \hline
  & $\sigma_{elas}$ &4.69E-4&4.84E-4&4.71E-4&6.70E-3\\ \hline
  & $\nu$ &1.06E-2 &9.31E-03&6.84E-3&1.22E-1\\ \hline
$^{238}$U &$\sigma_{f}$ &3.24E-2&3.23E-2&3.21E-2&4.61E-1\\ \hline
  & $\sigma_{n,\gamma}$ &1.37E-2&2.19E-2&1.70E-2&2.13E-1\\ \hline
  & $\sigma_{inelas}$ &6.70E-2&7.24E-2&6.93E-2&9.93E-1\\ \hline
  & $\sigma_{elas}$ &9.94E-2&1.03E-1&9.84E-2&1.42E-0\\ \hline
  & $\nu$ &2.64E-3&2.33E-3&1.71E-3&3.05E-2 \\ \hline
$^{239}$Pu &$\sigma_{f}$ &4.83E-1&3.43E-1&2.29E-1&6.74E-2 \\ \hline
  & $\sigma_{n,\gamma}$ &3.53E-2&7.93E-3&1.95E-3&8.98E-2\\ \hline
  & $\sigma_{inelas}$ &1.62E-4&1.68E-4&1.62E-4&2.34E-3\\ \hline
  & $\sigma_{elas}$ &2.73E-4&2.86E-4&2.74E-4&3.92E-3\\ \hline
  & $\nu$ &2.67E-3&2.35E-3&1.73E-3&3.09E-3\\ \hline
$^{241}$Pu &$\sigma_{f}$ &1.11E-1&1.362&1.11E-1&2.59E-2 \\ \hline
  & $\sigma_{n,\gamma}$ &1.99E-3 &2.56E-3&6.49E-4&1.91E-2\\ \hline
  & $\sigma_{inelas}$ &5.82E-5&5.87E-5&5.89E-5&8.33E-4\\ \hline
  & $\sigma_{elas}$ &3.32E-5&4.42E-5&3.16E-5&4.52E-4\\ \hline
  & $\nu$ &1.61E-3&1.42E-3&1.04E-3&1.86E-2\\ \hline
$^{16}O$ & $\sigma_{inelas}$ &5.78E-5 &5.94E-5&6.39E-5&8.49E-4\\ \hline
  & $\sigma_{elas}$ &6.64E-2&2.51E-2&6.56E-3&4.31E-1\\ \hline
H & $\sigma_{elas}$ &6.09E-2&7.08E-2&5.33E-2&8.99E-1\\ \hline
Total & & 5.51E-1 & 1.42E+0&4.15E-1& 2.07E+0 \\

\bottomrule

\end{tabular*}
\end{center}
\footnotesize
{$\sigma_{n,\gamma}$ represent capture cross section,$\sigma_{f}$ fission cross section, $\sigma_{elas}$  elastic scattering cross section,$\sigma_{inelas}$  inelastic scattering cross section, $\nu$ neutron number per fission}
\end{table}

\begin{figure}
\begin{center}
\includegraphics[width=9cm]{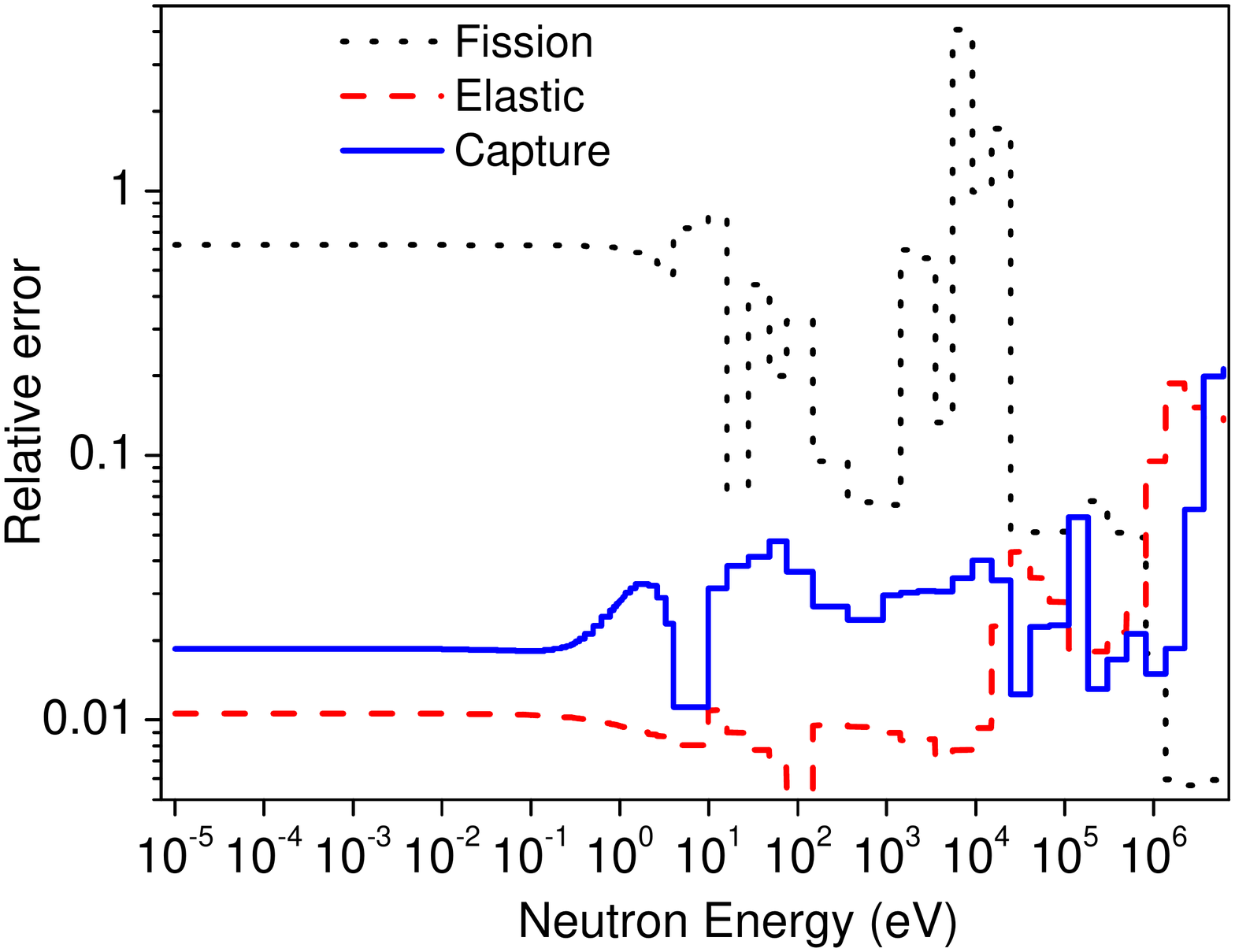}
\caption{Cross section relative error of $^{238}$U vs the neutron energy }
\label{235U_nonequlibrium}
\end{center}
\end{figure}

\begin{table}
\begin{center}
\caption{\label{umb}
Fission fraction uncertainties caused by atomic density change during burnup (\%)}
\footnotesize
\begin{tabular*}{140mm}{c@{\extracolsep{\fill}}ccccc}
\toprule
Isotopes~~  &$^{239}$Pu &$^{241}$Pu & $^{235}$U & $^{238}$U  \\
$^{235}$U &1.07E+00	&1.12E+00 & 5.36E-01& 1.93E-01\\ \hline
$^{238}$U &5.34E-02&6.03E-02&5.97E-02&8.24E-01\\ \hline
$^{239}$Pu &1.29E+00	& 5.56E-01 & 5.16E-01& 7.24E-03\\ \hline
$^{241}$Pu &7.17E-02&1.73E+00 & 7.10E-02 & 3.27E-03\\ \hline
Total	&1.68E+00&	2.14E+00& 7.50E-01& 8.46E-01 \\
\bottomrule
\end{tabular*}
\end{center}
\end{table}

\begin{table}
\begin{center}
\caption{\label{umm}
Fission fraction uncertainties caused by manufacturing parameters (\%)}
\footnotesize
\begin{tabular*}{140mm}{c@{\extracolsep{\fill}}ccccccc}
\toprule
Isotopes~~  &Clad &Pellet & Fuel & Gap& $^{235}$U  &Total \\
 & thickness& diameter & density & thickness & enrichment & & \\  \hline
$^{235}$U &2.12E-15 &5.37E-5 &2.82E-4&5.45E-5&9.03E-5&2.92E-4\\ \hline
$^{238}$U &1.01E-15&6.99E-4 &3.91E-3&7.21E-4&3.81E-5&4.03E-3\\ \hline
$^{239}$Pu &1.46E-15 &2.80E-5 &2.01E-4& 3.04E-5& 1.79E-4 &2.05E-4\\ \hline
$^{241}$Pu &1.13E-15 &5.6E-5 &2.98E-4&6.10E-5&1.87E-4 & 4.03E-3\\
\bottomrule
\end{tabular*}
\end{center}
\end{table}

\begin{table}
\begin{center}
\caption{\label{umt}
Total fission fraction uncertainties(\%)}
\footnotesize
\begin{tabular*}{140mm}{c@{\extracolsep{\fill}}ccccc}
\toprule
Operation state &$^{239}$Pu &$^{241}$Pu & $^{235}$U & $^{238}$U  \\ \hline
BOC &	1.80 &2.59&0.81&2.24 \\ \hline
MOC &	1.79 &2.58&0.83&2.24 \\ \hline
EOC &	1.77 &2.57&0.86&2.24 \\
\bottomrule
\end{tabular*}
\end{center}
\end{table}

\section{Conclusion}
Reactor simulation uncertainties were evaluated using atomic density comparison calculations from experiments reported previously . In this study, Monte Carlo-based sampling was used to evaluate the fission fraction uncertainties. It was found that fission cross section uncertainties are important uncertainty sources for $^{235}$U, $^{239}$Pu, and $^{241}$Pu, but for $^{238}$U, elastic and inelastic cross sections are important. Among the manufacturing parameters, the fuel density was the main source of uncertainty; however, the overall manufacturing parameter uncertainties were very small. The uncertainties induced by burnup were evaluated using the uncertainty of atomic density of the four isotopes. The total fission fraction uncertainties in reactor simulation at the MOC were 0.83\%, 2.24\%, 1.79\%, and 2.58\% for$^{235}$U, $^{238}$U, $^{239}$Pu, and $^{241}$Pu, respectively. The total fission fraction uncertainties were smaller than the previously obtained value of 5\%. These results are helpful for study of the reactor antineutrino anomaly and precise antineutrino spectrum measurement in the Daya Bay antineutrino experiment.

\section*{Acknowledgments}
The work was supported by National Natural Science Foundation of China (Grant No. 11390383) and the Fundamental Research Funds for the Central Universities(Grant No. 2018ZD10,2018MS044). We thank Bryan Schmidt from Liwen Bianji, Edanz Editing China (www.liwenbianji.cn/ac), for editing the English text of a draft of this manuscript.






\begin{thebibliography}{10}
%
\bibitem{DoubleChooz}Y.Abe et al. (Double Chooz Collaboration), Phys. Rev.
Lett. 2012, 108: 131801.
\bibitem{Dayabay}F.P.An et al. (Daya Bay Collaboration), Phys. Rev.
Lett. 2012, 108: 171803.
\bibitem{RENO}J.K.Ahn et al. (RENO Collaboration), Phys. Rev.
Lett. 2012, 108: 191802
\bibitem{safe}E. Christensen, P. Huber, P. Jaffke, and T. E. Shea, Phys. Rev.
Lett. 2014, 113: 042503.
\bibitem{spc1Dayabay}F.P. An et al. (Daya Bay Collaboration), Phys. Rev. Lett. 2017, 118: 099902
\bibitem{spc2Dayabay}F.P. An et al. (Daya Bay Collaboration),Chinese Physics C. 2017, 41(1):013002
\bibitem{spc2Chooz}Y. Abe et al. (Double Chooz Collaboration), Journal of High Energy Physics 2014,2014: 86.
\bibitem{a7}Th. A. Mueller et al., Phys. Rev. C 83, 054615 (2011).
\bibitem{a8} F. von Feilitzsch, A. A. Hahn, and K. Schreckenbach, Phys. Lett.
B 118, 162 (1982).
\bibitem{a9} K. Schreckenbach, G. Colvin,W. Gelletly, and F. von Feilitzsch,
Phys. Lett. B 160, 325 (1985).
\bibitem{a10} A. A. Hahn et al., Phys. Lett. B 218, 365 (1989).
\bibitem{spca1} D. A. Dwyer, T. J. Langford. Spectral Structure of Electron Antineutrinos from Nuclear Reactors, Phys. Rev. Lett. 114, 012502 (2015)
\bibitem{spca2} C.Buck, A.P.Collin, J.Haser, et al. Investigating the spectral anomaly with different reactor antineutrino experiments. PhysicsLettersB765(2017)159每162
\bibitem{spca3} A. A. Sonzogni, T. D. Johnson, E. A. McCutchan. Nuclear structure insights into reactor antineutrino spectra, PHYSICAL REVIEW C 91, 011301(R) (2015)
\bibitem{ZDjurcic2009}Z Djurcic, J A Detwiler, A Piepke, et al. Uncertainties in the anti-neutrino production at nuclear reactors, J.Phys.G: Nucl.Part.Phys. 2009, {\bf 36 }: 045002
\bibitem{Jones}C. L. Jones, A. Bernstein,et.al. Reactor simulation for antineutrino experiments using DRAGON and MURE, Phys. Rev. D,2012,{\bf 86}:012001
\bibitem{science}P.Girieud, L. Daudin, C. Garat, P. Marotte, S. Tarle, SCIENCE Version 2. The most recent capabilities of the Framatome 3D Nuclear Code Package, Proc. Int. Conf. ICONE 9, Nice Acropolis, France, April 8-12, 2001.

\bibitem{mpla1ma}X.B. Ma,F.Lu, L.Z.Wang,et.al. Uncertainty analysis of ssion fraction for reactor antineutrino experiments, Modern Physics Letters A,2016, Vol. 31, No. 20 : 1650120
\bibitem{scima2}X.B.Ma, R.M.Qiu,Y.X.Chen, et.al. New Monte Carlo-based method to evaluate fission fraction uncertainties for the reactor antineutrino experiment, Nuclear Physics A 2017,{\bf 958}:211每218
\bibitem{sandwich} D. G. CACUCI, Sensitivity and Uncertainty Analysis:
Theory, Vol. 1, Chapman and Hall/CRC, Boca Raton, Florida,2003.
\bibitem{sandwich1}Yousry Azmy.Enrico Sartori, Nuclear computational science A Century in Review, ISBN 978-90-481-3410-6, 2010, p219
\bibitem{njoy}R.E. MACFARLANEF, A.C. KAHLER, "Methods for Processing ENDF/B-VII with NJOY", Nuclear Data Sheets, 111, 12, 2739 (2010) (see http://t2.lanl.gov/codes/njoy99/).
\bibitem{b7} http://www.nndc.bnl.gov/endf/b7.1/
\bibitem{jendl} http://wwwndc.jaea.go.jp/jendl/j40/j40.html
\bibitem{suacl}JiaYi-Xu, Xu Bo-Ma,Fan Lu,et al. Nuclear Data and Fuel/Assembly Manufacturing Uncertainties Analysis and Preliminary Validation of SUACL, 2017 https://arxiv.org/abs/1704.06601
\bibitem{dragon} R.R.G. Marleau, A. Hebert and R. Roy, A User Guide for DRAGON, Technical Report, IGE-236 Rev. 1 (2001).
\bibitem{benchmark}K.Ivanov, M.Avramova, S.Kamerow,et al. Benchmarks for uncertainty analysis in modelling (UAM) for the design, operation and safety analysis of LWRs, Volume I: Specification and Support Data for Neutronics Cases (Phase I),NEA/NSC/DOC(2013)7
\bibitem{akioyamamoto} Akio Yamamoto, Kuniharu Kinoshita, Tomoaki Watanabe, et al. Uncertainty Quantification of LWR Core Characteristics Using Random Sampling Method, NUCLEAR SCIENCE AND ENGINEERING,2015, {\bf 181}: 1$-$15
\bibitem{phdfull}MATTHEW RYAN BALL, UNCERTAINTY IN LATTICE REACTOR PHYSICS CALCULATIONS, McMaster University, 2011
\bibitem{casmo4}Pusa M. Incorporation sensitivity and uncertainty analysis to a lattice physics code with application to CASMO-4, Annals of Nuclear Energy,2012, 40:153-162

\bibitem{TSUNAMI} B. T. Rearden, D. E. Mueller, S. M. Bowman, R. D. Busch, and S. J. Emerson, TSUNAMI Primer: A Primer for Sensitivity/Uncertainty Calculations with SCALE, ORNL/TM-2009/027, Oak Ridge National Laboratory, Oak Ridge, Tenn., January 2009
\bibitem{unicorn}Wan Cheng-hui,Cao Liang-zhi,Wu Hong-chuan,et al. Eigenvalue Uncertainty Analysis Based on Statistical Sampling Method, Atomic Energy Science and Technology, 2015,49,11
\bibitem{mw900}Guangdong Nuclear Power Training Center, Devices \& Systems of 900 MW PWR, Atomic Energy Press, ISBN7-5022-3171-4, 2005, pp.50$-$51.
\bibitem{arp} Alain,Hebert, Applied Reactor Physics, press internationales polytechnique,2009,p67\-79
\bibitem{cjd} C.J.Diez,O.Buss,A.Hoefer, et al. Comparison of nuclear data uncertainty propagation methodologies for PWR burn-up simulations. Annals of Nuclear Energy 77 (2015) 101每114, arXiv:1411.0834v1,2014.



\end{thebibliography}
\end{document}